\documentclass[a4paper,12pt]{article}
\usepackage[english]{babel}
\usepackage{amsmath}
\usepackage{epsfig}

\begin{document}

\textbf{Memory Effects in Turbulent Dynamo: Generation and
Propagation of Large Scale Magnetic Field}

\vskip 0.5cm

\centerline{Sergei Fedotov$^1$ , Alexey Ivanov$^{1, 2}$ and Andrey
Zubarev$^{1, 2}$}

\begin{center}
$^1$ Mathematical Department, UMIST - University of Manchester Institute
Science and Technology Manchester M60 1QD UK,

e-mail: Sergei.Fedotov@umist.ac.uk

$^2$ Department of Mathematical Physics, Ural State University, Lenin Av.,
51, 620083 Ekaterinburg, Russia.

e-mail: Alexey.Ivanov@usu.ru; Andrey.Zubarev@usu.ru
\end{center}

\

PACS numbers 47.65.+a, 95.30.Qd

Abstract.

We are concerned with large scale magnetic field dynamo generation
and propagation of magnetic fronts in turbulent electrically
conducting fluids. An effective equation for the large scale
magnetic field is developed here that takes into account the
finite correlation times of the turbulent flow. This equation
involves the memory integrals corresponding to the dynamo source
term describing the alpha-effect and turbulent transport of
magnetic field. We find that the memory effects can drastically
change the dynamo growth rate, in particular, non-local turbulent
transport might increase the growth rate several times compared to
the conventional gradient transport expression. Moreover, the
integral turbulent transport term leads to a large decrease of the
speed of magnetic front propagation.

\newpage

\bigskip

The problem of the generation and propagation of a magnetic field
in turbulent electrically conducting fluids is of fundamental
importance due to various applications in plasma physics,
astrophysics, geophysics, etc. \cite {Mof}-\cite{RShS}. It has
attracted much attention since the late 60s when it was realized
that the conditions for the occurrence of a large scale magnetic
field can be found by applying a simple technique involving the
mean helicity of the turbulence \cite{KrR}. This technique is
based on the effective macroscopic equation governing the large
scale magnetic field $\ \mathbf{B}$. The standard form of the mean
field dynamo equation is

\begin{equation}
\frac{\partial \mathbf{B}}{\partial t}=\textrm{rot}(\alpha
\mathbf{B})+\beta \Delta \mathbf{B}+\textrm{rot}[\textbf{u}\times
\textbf{B}]\,
\end{equation}
\ where $\ \mathbf{u}\ $ is the mean velocity field, $\ \alpha \ $
is the helicity and $\ \beta \ $ is the turbulent diffusivity.
This equation is a common starting point for analyzing the
generation of the large-scale magnetic field \cite{Mof}-
\cite{Nunez}. It has been also used for the analysis of the
propagation of magnetic fronts in spiral galaxies \cite{Ruzm}-
\cite{Soward}.

The main disadvantage of equation (1) is that it has been derived
for turbulent flow involving only two separated length scales for
the velocity field - the integral length scale and the small
turbulent scale \cite{KrR}. It is clear that the assumption of two
separated scales is rather unrealistic for fully developed
turbulent flow, which involves a continuous range of spatial and
temporal scales \cite{MY}. Thus, the purely local equation (1) is
applicable only under the assumption of a clear-cut separation
between macroscopic behavior of the averaged magnetic field and
the turbulent fluctuations at the ''microscopic level''.

It should be noted that the equation (1) is similar to the
convection-diffusion-reaction equations \cite{Mur, Br}. In fact,
it can be reduced to a famous
Fisher-Kolmogorov-Petrovskii-Piskunov (FKPP) equation \cite{RShS,
Ruzm}, which has become a basic mathematical tool in the theory of
propagating fronts travelling into the unstable state of the
reaction-diffusion systems. There has been an increased interest
in this topic, because of the large number of physical, chemical
and biological problems that can be treated in terms of the FKPP
equation (see, for example, \cite{Mur}-\cite{Had}). Recently,
there has been a tremendous activity to extend this analysis by
introducing more realistic description of the transport processes.
The main motivation for this is that the diffusion approximation
for transport admits an infinite speed of propagation. Due to this
nonphysical property of the diffusion solution, the FKPP equation
yields an overestimation of the propagation speed of travelling
fronts \cite {Had}-\cite{Eb}. We expect that a similar situation
might take place in magnetohydrodynamics. The mean-field dynamo
equation (1) also admits an infinite speed of a magnetic field
propagation. It is clearly a nonphysical property, because the
speed of magnetic field propagation can not exceed the velocity of
the largest eddy of turbulent flow. The origin of this
contradiction lies in the $\ \delta -$correlated-in-time
approximations for the turbulent velocity field (see below). In
reality these correlations have finite times of relaxation, and
what is more these might be of the same order as the
characteristic times for the growth rate of the large scale
magnetic field since the physical origin of these correlations and
the magnetic field generation are the same, namely, the turbulent
fluctuations.

It is the aim of this Letter to extend mean field dynamo equation
(1) to the case when long range in time correlations of turbulent
flow are taken into account and find out how the non-local in time
effect might influence the critical conditions for the generation
of magnetic field and its spatial propagation.

The most general phenomenological formulation of the dynamo
problem that is considered here is represented by the equations
for the average magnetic field $\mathbf{\ B}$ \cite{KrR}:
\begin{equation}
\frac{\partial \mathbf{B}}{\partial t}=\textrm{rot}\mathcal{E}+\textrm{rot}[%
\textbf{u}\times \textbf{B}]\ ,\;\mathcal{E}=\left\langle \mathbf{u}%
^{\prime }\times \mathbf{B}^{\prime }\right\rangle ,
\end{equation}
where primes denote the turbulent fluctuations and the angular
brackets denote an ensemble averaging over the turbulent
pulsations. The main closure problem here is to express the
turbulent electromotive force $\ \mathcal{E}\ $\ in terms of the
average field $\mathbf{\ B.}$ The classical expression
leading to (1) is $\mathcal{E}=\alpha \ \mathbf{B}\ -\beta \ $rot$\mathbf{B}%
,\ $where $\alpha \ $\textbf{\ }and $\beta \ $are the statistical
characteristics of the turbulent velocity $\mathbf{u}^{\prime }.$
However, under the assumptions of infinite conductivity and weak
turbulence the electromotive force $\ \mathcal{E}$  can be written
as \cite{KrR}:
\begin{equation}
\mathcal{E}(\mathbf{x},t)=-\frac{1}{3}\int_{-\infty
}^{t}\left\langle
\mathbf{u}^{\prime }(\mathbf{x},t)\cdot \ \textrm{rot}\,\mathbf{u}^{\prime }(%
\mathbf{x},s)\right\rangle \mathbf{B}(\mathbf{x},s)\ ds-
\end{equation}
\[
-\frac{1}{3}\int_{-\infty }^{t}\left\langle \mathbf{u}^{\prime }(\mathbf{x}%
,t)\cdot \mathbf{u}^{\prime }(\mathbf{x},s)\right\rangle \textrm{rot}\mathbf{B}%
(\mathbf{x},s)\ ds\ .
\]
\

The local mean field equation (1) can be derived from (3) under
the assumptions that the correlations appearing in (3) are
approximated by the delta-functions in time: $\ \left\langle
\mathbf{u}^{\prime }(\mathbf{x},t)\cdot
\textrm{rot}\,\mathbf{u}^{\prime }(\mathbf{x},s \right\rangle
=-3\alpha
\ \delta (t-s),\ $\ $\left\langle \mathbf{u}^{\prime }(\mathbf{x}%
,t)\cdot \mathbf{u}^{\prime }(\mathbf{x},s)\right\rangle = 3\beta
\ \delta (t-s)$. However, as was mentioned in \cite{ZRS} (p. 136)
''... the assumption of instantaneous correlations seems to be a
serious restriction to the theory...'', since in real turbulence
the characteristic times of these correlations are finite. It
follows from (3) that in the case of finite correlation times the
electromotive force should contain the integrals over the history
of $\ \mathbf{B}$ and rot$\mathbf{B.}$

In this Letter we suggest the following general form for the
electromotive force $\mathcal{E}$ in the limit of infinite
conductivity

\begin{equation}
\mathcal{E}=\alpha (\mathbf{x})\int_{-\infty }^{t}G_{\alpha }\left( \frac{t-s%
}{\tau _{\alpha }}\right) \ \mathbf{B}(\mathbf{x},s)\ ds-\beta (\mathbf{x}%
)\int_{-\infty }^{t}G_{\beta }\left( \frac{t-s}{\tau _{\beta
}}\right) \ \textrm{rot}\mathbf{B}(\mathbf{x},s)\ ds\ ,
\end{equation}
\ where $G_{\alpha }\left( y\right) $ and $G_{\beta }\left(
y\right) $ are positive, decreasing functions that tend to zero as
$y\rightarrow \infty .$ The parameters $\ \tau _{\alpha }\ $ and
$\ \tau _{\beta }\ $ control the time correlations in the random
velocity field at a fixed space position. By inserting (4) into
(3) one can get the non-local in time mean field dynamo equation:

\begin {equation}
\frac{\partial \mathbf{B}}{\partial t}= \int_{-\infty
}^{t}G_{\alpha }\left ( \frac{t-s}{\tau_\alpha} \right ) \
\textrm{rot}\left[ \alpha \textbf{B}(s)\right] \ ds -
\end{equation}
$$
- \int_{-\infty }^{t}G_{\beta }\left( \frac{t-s}{\tau_\beta}
\right ) \ \textrm{rot} \left [ \beta \ \textrm{rot} \mathbf{B}(s)
\right ] \ d s +\textrm{rot}[ \textbf{u}\times \textbf{B}]
 \ . $$
It should be noted that this equation is valid only in the limit
of infinitely large conductivity. This case is of specific
interest for magnetohydrodynamics of plasma and many problems of
astrophysics and geophysics \cite{KrR}. Many different
constitutive models might arise from different choices of
$G_{\alpha }\left( y\right) $ and $G_{\beta }\left( y\right) .$
The equation (1) can be considered as a limiting case of (5) when
$\ G_{\alpha ,\beta }(t-s)=\delta (t-s)$ $\left( \tau _{\alpha
,\beta }\rightarrow 0\right) .\ $ The transport memory kernel $\
G_{\beta }\ $ ensures the finite velocity of propagation of a
magnetic field, which is determined by the rate of turbulent
pulsations \cite{MY}. The relaxation
time $\tau _{\beta }$ can be determined from the root mean square velocity $%
u_{rms}=\sqrt{\beta /\tau _{\beta }}$ of turbulent fluctuations.
The integral kernel $\ G_{\alpha }\ $ is used to express the fact
that the dynamo growth of a magnetic field at a time $\ t\ $ in
the local vicinity of the spatial point $\ \textbf{x}\ $ is
determined by the past values of rot$\textbf{B}$.

Although there has been a large number of studies of the
generation of macroscopic magnetic fields, the implications of
finite correlations for the generation and propagation of the
large scale magnetic field have not been studied. The implications
of our results for dynamo theory are twofold. First, the
non-locality in time greatly influences the dynamo growth rate of
a magnetic field, however, the generation conditions are not
sensitive to the addition of the memory effect due to non-zero
time correlations: they are robust to the addition of non-local
terms for the mean field dynamo equation (5). Secondly, the memory
effects introduce drastic changes to the front dynamics of a
magnetic field.

In what follows, we study the influence of the memory effects on
the dynamo generation and propagation by using an important
example of a thin turbulent slab of thickness $\ 2h\ $ and radius
$\ R\ $ ($R\gg h$), which rotates with the angular velocity $\
\omega (r)$. This is a standard model for disc-like galaxies. We
neglect the effects of compressibility, diamagnetism and
deviations from the axial symmetry. We use the standard
approximation of constant turbulent diffusion coefficient $\ \beta
$ \cite{ZRS, RShS, Ruzm, Moss, Soward}. We restrict our analysis
to the kinematical aspects of the problem, neglecting the
influence of the magnetic field on the turbulent flow, i.e. the
dependencies of $\ \alpha ,\beta ,\tau _{\alpha }\ $ and $\ \tau
_{\beta }\ $ on the magnetic field $\ B$. With these
simplifications, the equations for the mean axisymmetric magnetic
field in polar cylindrical coordinates ($r,\varphi ,z$) with $\
z$-axis coincident with the rotation axis follow from the basic
equation (5):

\begin{equation}
\frac{\partial B_r}{\partial t}= - \int_{-\infty }^{t}G_{\alpha }\left (
\frac{t-s}{\tau_\alpha} \right ) \frac{\partial}{\partial z} (\alpha
B_\varphi)\ d s +
\end{equation}
\[
+ \beta \int_{-\infty }^{t}G_{\beta }\left( \frac{t-s}{\tau_\beta} \right )
\left\{ \frac{\partial^2 B_r}{ \partial z^2} + \frac{\partial}{\partial r}
\left [ \frac{1}{r} \frac{ \partial}{\partial r} \left ( r B_r \right )
\right ] \right\} d s \ ,
\]

\begin{equation}
\frac{\partial B_\varphi}{\partial t}= g \ B_r + \int_{-\infty
}^{t}G_{\alpha }\left ( \frac{t-s}{\tau_\alpha} \right ) \frac{ \partial}{%
\partial z} (\alpha B_r) \ d s +
\end{equation}
\[
+ \beta \int_{-\infty }^{t}G_{\beta }\left( \frac{t-s}{\tau_\beta} \right )
\left\{ \frac{ \partial^2 B_\varphi}{\partial z^2} + \frac{\partial}{%
\partial r} \left [ \frac{1}{r} \frac{\partial}{\partial r} \left ( r
B_\varphi \right
) \right ] \right\} \ d s \ .
\]
\

Here $\ g=rd\omega /dr\ $ is the measure of differential rotation,
and we are interested only in $\ B_{r}\ $ and $\ B_{\varphi }\ $
components of a magnetic field, because $\ B_{z}/B_{r,\varphi
}=O(h/R)\ $ \cite{ZRS, RShS}. These components obey the vacuum
boundary conditions on the thin disc surfaces \cite{ZRS, RShS}: \
$B_{r,\varphi }(z=\pm h)=0$. The main goal of our analysis is to
find the local growth rate of the magnetic field and its spatial
propagation, taking into account the memory effects. Thus,
considering the field generation in the usual way \cite{ZRS, RShS,
Ruzm, Moss}, we neglect first the radial derivatives in equations
(6), (7). We represent the components of the magnetic field as
follows $B_{r,\varphi }(z,t)=b_{r,\varphi }(z)\exp (\gamma \ t),$
where $b_{r,\varphi }(z)$ have to be found from the eigenvalue
problem

\begin{equation}
\left( \tilde{\gamma}+\frac{\partial ^{2}}{\partial z^{2}}\right) b_{r}=-%
\tilde{R}_{\alpha }\frac{\partial (\alpha b_{\varphi })}{\partial z}\ ,\ \ \
\left( \tilde{\gamma}+\frac{\partial ^{2}}{\partial z^{2}}\right) b_{\varphi
}=\tilde{R}_{w}b_{r}+\tilde{R}_{\alpha }\frac{\partial (\alpha b_{r})}{%
\partial z}\ ,\ b_{r,\varphi }(z=\pm 1)=0\ ,
\end{equation}
\ where $\tilde{\gamma} = \gamma / f_\beta ( \gamma T_\beta) \ , \
\ \ \tilde{R}_w = R_w / f_\beta ( \gamma T_\beta) \ , \ \ \
\tilde{R}_\alpha = R_\alpha f_\alpha ( \gamma T_\alpha / R_\alpha)
/ f_\beta ( \gamma T_\beta) \ , f_{\alpha, \beta} ( \gamma
\theta_{\alpha, \beta} )= \int_0^\infty G_{\alpha, \beta} ( \xi /
\theta_{\alpha, \beta} ) \exp(- \gamma \xi) \ d \xi \ , \ \ \
\theta_\alpha = T_\alpha / R_\alpha \ , \ \ \ \theta_\beta =
T_\beta \ . $ Here we use the dimensionless variables $\
z\rightarrow z/h,t\rightarrow \beta t/h^{2},\alpha \rightarrow
\alpha _{0}\alpha (z)$, and dimensional parameters $\ T_{\alpha
}=\alpha _{0}\tau _{\alpha }/h,T_{\beta }=\beta \tau _{\beta
}/h^{2},R_{\alpha }=\alpha _{0}h/\beta ,R_{w}=gh^{2}/\beta $. The
parameter $\ \gamma \ $ describes the growth rate of the magnetic
field. The eigenvalue problem (8) reduces to the well-known form
\cite{ZRS, RShS, Ruzm,
Moss} in the case when $\ T_{\alpha ,\beta }=0$. Otherwise, the eigenvalue $%
\ \gamma \ $ becomes a function of these correlation times, and
this function is determined from the problem (8) using the
renormalized parameters $\ \tilde{\gamma},\tilde{R}_{\alpha
},\tilde{R}_{w}$.

The problem (8) coincides with the generation equations in the
local mean-field dynamo theory \cite{ZRS, RShS, Ruzm, Moss}. The
only difference is that the renormalized parameters $\
\tilde{\gamma},\tilde{R_{\alpha }}$ and $\tilde{R_{w}}\ $ are
dependent on the increment $\ \gamma \ $ of the time growth of the
magnetic field. This means that the generation equations (8) are
universal and do not depend on the specific choice of the memory
kernels $\ G_{\alpha ,\beta }$. The main physical conclusion from
this universality is that since physically meaningful memory
kernels must satisfy the normalization condition $\ f_{\alpha
,\beta }(\gamma =0)=1$,
then the threshold combinations of dynamo parameters $\ R_{\alpha }\ $ and $%
\ R_{\beta }\ $ at a given $\ \alpha (z)$, providing the critical
point of instability ($\textrm{Re} \gamma =0$), are the same for
all memory kernels. These critical parameters can be determined
from the local mean-field dynamo theory. To determine the growth
rate $\ \gamma \ $ in the generation region, the
eigenvalue problem (8) should be solved as $\ \tilde{\gamma}=\tilde{\gamma}(%
\tilde{R}_{\alpha },\tilde{R}_{w})$. This relation with the
definitions of the renormalized parameters $\
\tilde{\gamma},\tilde{R}_{\alpha },\tilde{R}_{w}\ $ leads to a
transcendental equation for $\ \gamma $. The specific dependence
of $\ \gamma \ $ on the relaxation times $\ T_{\alpha ,\beta }\ $
and dynamo parameters $\ R_{\alpha ,w}\ $ has to be determined by
the specific forms of the memory kernels $\ G_{\alpha ,\beta }$.

Let us illustrate the general results by using the important
example of $\ \alpha \omega $ dynamo ($R_{\alpha }\ll |R_{w}|$),
which is of specific interest for astrophysical problems
\cite{KrR}-\cite{RShS}, \cite{Ruzm}-\cite {Soward}. We use the
exponential relaxation form
\begin{equation}
G_{\alpha ,\beta }(y)=\exp \left( - y /\tau _{\alpha ,\beta
}\right) /\tau _{\alpha ,\beta }\ .
\end{equation}
\
The solution of the eigenvalue problem (8) depends on the form
of the function $\ \alpha ( z ) $. The $\ \alpha \omega \ $
approximation requires the anti-symmetric character $\ \alpha ( z
) = - \alpha ( - z ) $. Following the asymptotic method of
solution of the eigenvalue problem (8) \cite{ZRS, RShS, Moss}, for
the model case $\ \alpha ( z ) = z $, the increment of maximal
growth satisfies the algebraic equation:

\begin{equation}
\gamma (1+\gamma T_{\beta })=-\frac{\pi ^{2}}{4}+\sqrt{D}\frac{1+\gamma
T_{\beta }}{\sqrt{1+\gamma T_{\alpha }/R_{\alpha }}}\ ,\ \ \ D=-R_{\alpha
}R_{w}\ .
\end{equation}
\

The growth rate $\ \gamma \ $ is shown in Fig. 1 as a function of
dimensionless correlation times $\ T_{\alpha }\ $ and $\ T_{\beta
}$. Clearly, the non-local terms corresponding to the turbulent
helicity and the turbulent transport influence the growth rate $\
\gamma \ $ in opposing way. The turbulent $\alpha $-source
non-locality leads to a slower field growth (curve 1) due to the
physical nature of the process: the integral representation of the
$\alpha $-term in basic equations (4), (5) introduces
the time memory of excitation. The non-locality of turbulent $\beta $%
-transport reduces the energy loss out of the disc and, hence,
results in an increase of the field growth rate (curve 2).
Moreover, for dynamo numbers $\ D$, close to the critical value $\
D_{cr}$, the influence of the transport non-locality becomes very
significant, the growth rate $\gamma $ increases drastically with
$\ T_{\beta }$. It is worth noting that the critical value $\
D_{cr}\simeq \pi ^{4}/16\ $ of the dynamo number $\ D$, at which
the generation commences ($\gamma =0$), does not depend on the
memory effects; this result is in agreement with the general
conclusion made above.

Now we are in a position to discuss the problem of magnetic front
propagation. If the dynamo excitation takes place within a certain radius
(say, $\ r<r_{0}$) then the magnetic field can propagate in the form of
travelling fronts \cite{Moss}. According to the local dynamo theory the
speed $\ v\ $ of a propagating magnetic front is proportional to $\ \sqrt{%
\beta \gamma }\ $ due to the FKPP type of the mean field equation
(1). The problem is that in the local mean-field description of
magnetic fields, turbulent diffusion induces an infinitely long
Gaussian tail ahead of magnetic front which leads to the
overestimation of the propagation rate. A qualitatively different
situation appears when the memory effects are taken into account.
This means that the speed $\ v\ $ can not exceed the maximum
possible velocity $\sqrt{\beta /\tau _{\beta }}$ however large the
growth rate $\gamma $ is$.$

Now let us turn to the problem of front propagation in the case
when the dynamo equation with memory (5) is considered in the thin
disk approximation. Assume that the initial distribution of
magnetic field satisfies: $\ B=B_{0}=const\ $ if $\ r<r_{0}\ $ and
$\ B=0\ $ otherwise. The main quantity of interest is the speed $\
v\ $ at which magnetic field propagates in the form of a
self-similar wave for the
large values of $\ r\ $ and $\ t\ .$ In the large-distance limit ($%
r\rightarrow \infty )$ the radial Laplace operator $\ \frac{\partial }{%
\partial r}(\frac{1}{r}\frac{\partial }{\partial r}r)\ $ in equations (6)
and (7) can be approximated by the second derivative $\ \frac{\partial ^{2}}{%
\partial r^{2}}\ $ and we may find the solution of these equations in the form
of Fourier modes $\ B_{r,\varphi }(r,z,t)=b_{r,\varphi }(z)\exp
(\gamma \ t)\exp (ikr)$. Substitution of this expression into
equations (6) and (7) leads to the problem for eigenfunctions $\
b_{r,\varphi }\ $ \cite{ZRS},\cite {RShS}, from which \ one can
find the equation for the exponent $\ \gamma (k)$ as a function of
the wave number $\ k$. The general theory of front propagation in
non-local reaction-diffusion media \cite{Fed, Eb}, based on the
saddle-point method of calculation of the inverse Fourier
integrals, leads to the following equation for the propagating
velocity $\ \ v=\gamma (\lambda )/\lambda \ $, where $\ \lambda
=ik\ $ satisfies the equation $\ \gamma (\lambda )/\lambda
=d\gamma (\lambda )/d\lambda $. We found that the front velocity
$\ v\ $ is a monotonically decreasing function of both correlation
times $\ \tau _{\alpha ,\beta }\ $. In particular, it depends
weakly on $\ \tau _{\alpha }\ $ (see Fig. 2) , but decreases
significantly (up to 2-3 times) with growing $\ \tau _{\beta }\ $.
Therefore, the use of FKPP-like estimation for the travelling wave
velocity $\ v\sim \sqrt{\beta \gamma }\ $ for systems with memory
leads to significant overestimations. This general conclusion is
of great importance not only for magnetic dynamo front propagation
but for a wide class of excitable media.

Basically, we have extended the classical mean field dynamo theory
to the case when the memory effects are taken into account. We
have suggested the effective equation for the large scale magnetic
field that takes into account the finite correlation times of the
turbulent flow. This equation involves memory integrals
corresponding to the dynamo source term describing the
alpha-effect and turbulent transport of magnetic field. We have
found that memory effects can drastically change the dynamo growth
rate, in particular,  the finite turbulent transport involving
memory might increase the growth rate several times. We have also
found that memory effects lead to the essential decrease of the
speed of magnetic front propagation.

The authors thank David Moss for interesting discussions. This
work was carried out under EPSRC Grant GR/M72241 and with the
financial support of the CRDF Award No. REC-005. One of the
authors (AI) was partly supported by the RME Grant No.
E00-3.2-210.

\newpage

\begin{figure}[tb]
\epsffile{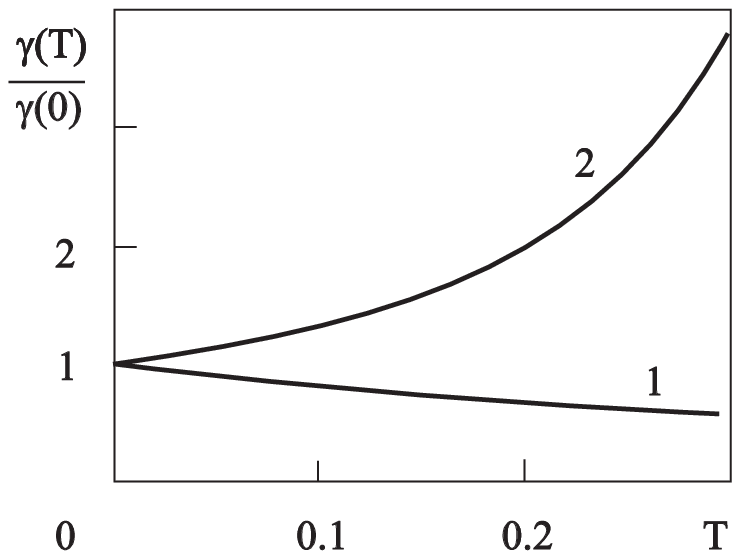} \caption{ Dependencies of the dimensionless
growth rate $\gamma $ on the turbulent $\alpha$-source relaxation
time $T_\alpha $ at $T_\beta = 0 $ (curve 1) as well as on the
transport time $T_\beta$ when $T_\alpha=0$ (curve 2). The memory
kernels $G_{\alpha,\beta}$ are chosen to have form (9); the
dimensionless dynamo parameters of the system are: $R_\alpha=1 $,
$R_w=6.1$.}
\label{fig1}
\end{figure}

\begin{figure}[tb]
\epsffile{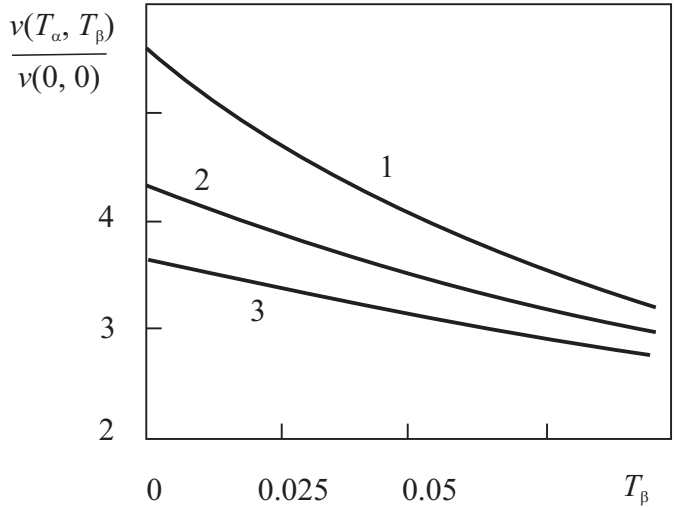} \caption{Dependence of the dimensionless front velocity ${\widetilde{v}%
=v(T_\alpha, T_\beta)/ v (0, 0)}$ on the relaxation time $T_\beta$ at $%
T_\alpha = 0$ (curve 1); 0.3 (curve 2) and 0.9 (curve 3). The
dynamo parameters are $R_\alpha = 2$ and $R_w = - 40 $ }
\label{fig2}
\end{figure}

\end{document}